\documentclass[journal]{IEEEtran}
\usepackage{graphicx}
\usepackage{amsmath}

\begin{document}

\title{An Accurate Approximation to the Distribution of the Sum of Equally Correlated Nakagami-$m$ Envelopes and its Application in Equal Gain Diversity Receivers}

\author{Zoran~Hadzi-Velkov, Nikola Zlatanov, and George K. Karagiannidis \\%
\vspace{-9mm}
\thanks{Accepted at IEEE ICC 2009}
\thanks{Z. Hadzi-Velkov and N. Zlatanov are with the Faculty of Electrical Engineering and Information Technologies, Ss. Cyril and Methodius University, Skopje, Email: zoranhv@feit.ukim.edu.mk,
nzlatanov@manu.edu.mk}
\thanks{G. K. Karagiannidis is with the Department of Electrical and Computer Engineering, Aristotle University of Thessaloniki, Thessaloniki, \qquad \qquad \qquad Email: geokarag@auth.gr}
}

\markboth{}{Shell \MakeLowercase{\textit{et al.}}: Bare Demo of
IEEEtran.cls for Journals} \maketitle

\begin{abstract}
We present a novel and  accurate approximation for the
distribution of the sum of equally correlated Nakagami-$m$
variates. Ascertaining on this result we study the performance of
Equal Gain Combining (EGC) receivers, operating over equally
correlating fading channels. Numerical results and simulations
show the accuracy of the proposed approximation and the validity
of the mathematical analysis.
\end{abstract}


\section{Introduction}

\PARstart{T}{he}  knowledge of the statistics of the sums of
multiple signals's envelopes is important in the analytical
performance evaluation, such as that of equal gain combining (EGC)
systems. However, the evaluation of the probability distribution
function (PDF) and the cumulative distribution function (CDF) of
these sums can be rather cumbersome even for the statistically
independent Nakagami-$m$ or Rayleigh fading channels
\cite{2}-\cite{7}. An inf\-inite series technique for computing
the PDF of a sum of independent random variables (RVs) was derived
in \cite{2}. Applying this technique, the error rate performance
of EGC systems under Nakagami fading was presented in \cite{3},
whereas, in \cite{4}, the problem was analyzed in frequency domain
in terms of semi-analytical expressions with inf\-inite integrals.
Other two studies on EGC diversity in Nakagami fading that use
numerical integration over Gil-Palaez single inf\-inite integral
and Hermite quadrature over double f\-inite-inf\-inite integral
are presented in \cite{5} and \cite{6}, respectively. Closed form
solutions for some modulation schemes are also obtained for dual
and triple diversity under Rayleigh fading \cite{2}, \cite{5}. All
above mentioned works assumed independent fading channels.

However, in real-life applications, fading among diversity
branches is correlated, which renders the analytical analysis
under correlated Nakagami fading with a particular practical
interest. Since the joint PDF of multiple correlated fading
branches is not known, the published results for EGC diversity in
correlated fading channels deal primarily with the dual branch
case \cite{7}-\cite{9}, where error probabilities for binary and
QAM signals over correlated Rayleigh channels are expressed in
form of inf\-inite series.

Only several papers address EGC in correlated fading with multiple
order diversity. In \cite{10}, EGC performance was determined by
approximating the moment generating function (MGF) of its output
SNR, where the moments are determined exactly for exponentially
correlated Nakagami channels in terms of multi-fold inf\-inite
series. A completely novel approach for performance analysis of
diversity combiners in equally correlated fading channels was
proposed in \cite{11}, where the equally correlated Rayleigh
fading channels are transformed into a set of conditionally
independent Rician RVs. Based on this technique, the authors in
\cite{12} derive the moments of the EGC output SNR in equally
correlated Nakagami channels in terms of the Lauricella
hypergeometric function, and then uses them to evaluate the EGC
performance measures, such as outage probability (as inf\-inite
series) and error probability (using Gaussian quadrature with
weights and abscissas computed by solving sets of nonlinear
equations).

All of the above approaches yield to results that are somewhat
complex, not expressed in closed form, and require computation of
inf\-inite series, all of which is attributed to the inherent
intricacy of the exact sum statistics. This intricacy can be
circumvented by searching for suitable highly accurate
approximations for a sum of arbitrary number of Nakagami RVs.
Various simple and accurate approximations to the PDF of sum of
independent Rayleigh, Rice and Nakagami RVs are proposed in
\cite{13}-\cite{15}, which then are used for analytical EGC
performance evaluation. \cite{15} uses the moment matching method
to arrive at the required approximation.

In this paper, we use the moment matching method to obtain highly
accurate closed form PDF approximation for the sum of arbitrary
number of non-identical equally correlated Nakagami RVs with
arbitrary mean powers. We then apply this approximation to
eff\-iciently estimate the performance of EGC systems by avoiding
many complex numerical calculations inherent for the methods in
abovementioned previous works. Even approximate closed form
expressions allow one to gain insight into system performance by
considering, for example, large SNR or small SNR behaviors.


\section{An Accurate Approximation to the Sum of Equally Correlated Nakagami-$m$ Envelopes}

Let $Z$  be a sum of $L$ non-identical equally correlated
Nakagami-$m$ RVs, $Z_1$, $Z_2$,\,...\,, $Z_L$,
\begin{equation}\label{1}
Z=\sum_{k=1}^L Z_k ,
\end{equation}

The PDF of each envelope $Z_k$, $1\leq k\leq L$ is given by \cite{1}
\begin{equation}\label{2}
f_{Z_k}(z)=\left(\frac{m_z}{\Omega_k}\right)^{m_z} \frac{2 z^{2
m_z-1}}{\Gamma(m_z)}
\exp\left(-\frac{m_z}{\Omega_k}z^2\right),\qquad z\geq 0
\end{equation}
having an arbitrary second moment $E[Z_k]=\Omega_k$, $1\leq k\leq
L$, same fading parameter $m_z$  (assumed to be positive integer)
and same envelope correlation coeff\-icient between each pair of
RVs
\begin{equation}\label{3}
\rho_Z = \frac{{\rm{cov}} (Z_i^2,Z_j^2)}{\sqrt{{\rm {var}}(Z_i^2)
{\rm {var}}(Z_j^2)}},\qquad i\neq j
\end{equation}
where $E[\cdot]$,  $\rm {cov} (\cdot,\cdot)$   and $\rm
{var}(\cdot)$ denote expectation, covariance and variance,
respectively.

We propose the unknown PDF of be approximated by the PDF of an
equivalent RV def\-ined as
\begin{equation}\label{4}
R=\sqrt{\sum_{k=1}^L R_k^2} ,
\end{equation}
where $R_k$, $1\leq k\leq L$, represent a different set of $L$
identical equally correlated Nakagami RVs with equal average
powers, $E[R_k]=\Omega_R$, equal fading parameters $m_R$  and
equal correlation coeff\-icient $\rho_R$. Additionally, it is
assumed that
\begin{equation}
\rho_R = \rho_Z = \rho \,.
\end{equation}
Both the MGF and the PDF of $R^2$ had been determined in closed
form as [16, Eqs. (42a) and (36)] \setlength{\arraycolsep}{0.0em}
\begin{eqnarray}\label{5}
M_{R^2}(s)=E[e^{sR^2}]=\left(1-s\frac{\Omega_R(1+(L-1)\sqrt{\rho})}{m_R}
\right)^{-m_R} \nonumber\\
\times \left(1-s\frac{\Omega_R(1-\sqrt{\rho})}{m_R}
\right)^{-m_R(L-1)} \,, \qquad
\end{eqnarray}
and
\begin{eqnarray}\label{6}
f_{R^2}(r)=\left(\frac{m_R}{\Omega_R}\right)^{m_R L} \qquad \qquad \qquad \qquad \qquad \qquad \qquad \quad \nonumber\\
\times \, \frac{r^{m_R L-1}}{\Gamma(m_R L)(1-\sqrt{\rho})^{m_R(L-1)}(1+(L-1)\sqrt{\rho})^{m_R} } \qquad \quad \nonumber\\
\times \, \exp\left(-\frac{m_R r}{(1-\sqrt{\rho})\Omega_R}\right) \qquad \qquad \qquad \qquad \qquad \qquad \quad \nonumber\\
\times \, {}_1F_1\left(m_R,m_R L,\frac{\sqrt{\rho} m_R L r}
{(1-\sqrt{\rho})(1+(L-1)\sqrt{\rho})\Omega_R }\right) \,, \qquad
\end{eqnarray}
respectively, where ${}_1F_1(\cdot,\cdot,\cdot)$  is the Kummer
conf\-luent hypergeometric function [17, Eq. (9.210)]. The PDF of
$R$ is determined by simple transformation of RVs, $f_R(r)=2r
f_{R^2}(r^2)$, which yields to
\begin{eqnarray}\label{7}
f_{R}(r)=\left(\frac{m_R}{\Omega_R}\right)^{m_R L} \qquad \qquad \qquad \qquad \qquad \qquad \qquad \quad \nonumber\\
\times \, \frac{2r^{2m_R L-1}}{\Gamma(m_R L)(1-\sqrt{\rho})^{m_R(L-1)}(1+(L-1)\sqrt{\rho})^{m_R} } \qquad \quad \nonumber\\
\times \, \exp\left(-\frac{m_R r^2}{(1-\sqrt{\rho})\Omega_R}\right) \qquad \qquad \qquad \qquad \qquad \qquad \quad \, \nonumber\\
\times \, {}_1F_1\left(m_R,m_R L,\frac{\sqrt{\rho} m_R L r^2}
{(1-\sqrt{\rho})(1+(L-1)\sqrt{\rho})\Omega_R }\right) \,. \qquad
\end{eqnarray}
One now needs to determine $\Omega_R$  and $m_R$  so that
(\ref{7}) be an accurate approximation of the PDF of $Z$ def\-ined
by (\ref{1}). For this, we apply the moment matching method by
respectively matching the second and fourth moments of RVs $Z$ and
$R$:
\begin{eqnarray}\label{8}
E[Z^2]& \,= \,&E[R^2] \,,\\
E[Z^4]& \,= \,&E[R^4] \,.
\end{eqnarray}
The second and the fourth moments of $R$ are determined straightforwardly by using the MGF (\ref{5}) and applying the moment theorem, i.e.,
\begin{eqnarray}\label{9}
E[R^2]&=&\frac{dM_{R^2}(s)}{ds}\big{|}_{s=0}=L\Omega_R \,,\\
E[R^4]&=&\frac{d^2M_{R^2}(s)}{ds^2}\big{|}_{s=0}=\frac{L\Omega_R}{m_R}(1-\rho+L(m_R+\rho)) \,.\nonumber\\
\end{eqnarray}
The second and fourth moments of $Z$ are determined by applying the multinomial theorem and the results presented
in [10, Eq. (21)], [12, Eq. (43)] and Appendix A, yielding
\begin{eqnarray}\label{10}
E[Z^2] = \sum_{k=1}^L\Omega_k+2\frac{\Gamma^2(m_z+1/2)}{m_z
\Gamma^2(m_z)}
\left(\sum_{i=1}^L\sum_{j=i+1}^L\sqrt{\Omega_i\Omega_j}\right)\nonumber\\
\times \, {}_2F_1\left(-\frac12,-\frac12;m_z;\rho\right) \,,
\qquad \qquad
\end{eqnarray}
and
\begin{eqnarray}\label{11}
E[Z^4]&=&\left(\frac{1-\sqrt\rho}{m_z}\right)^2\left[W(4)
\sum_{k=1}^L\Omega_k^2+6 W(2,2)\right.\nonumber\\
&\times&\sum_{i=1}^L\sum_{j=i+1}^L\Omega_i\Omega_j +4 W(3,1)
\sum_{i=1}^L\sum_{j=i+1}^L\left(\sqrt{\Omega_i^3\Omega_j}\right.\nonumber\\
&+&\left.\sqrt{\Omega_i\Omega_j^3}\right) +12 W(2,1,1)
\sum_{m=1}^L\sum_{i=m+1}^L\sum_{j=i+1}^L\nonumber\\
&&\left(\sqrt{\Omega_m^2\Omega_i\Omega_j}+
\sqrt{\Omega_m\Omega_i^2\Omega_j}+\sqrt{\Omega_m\Omega_i\Omega_j^2}\right)+24\nonumber\\
&\times& \left.W(1,1,1,1)\sum_{m=1}^L\sum_{n=m+1}^L\sum_{i=n+1}^L\sum_{j=i+1}^L
\sqrt{\Omega_m\Omega_n\Omega_i\Omega_j}\right]\nonumber\\
\end{eqnarray}
where
\begin{eqnarray}\label{12}
W(k_1,\cdots,k_N)=\left(\prod_{j=1}^N\frac{\Gamma(m_z+k_j/2)}{\Gamma(m_z)}\right)
\frac{1}{\Gamma(m_z)} \qquad \qquad \, \nonumber\\
\times \int_0^\infty u^{m_z-1}e^{-u}\prod_{j=1}^N
{}_1F_1\left(-\frac{k_j}{2},m_z;-\frac{u\sqrt\rho}{1-\sqrt\rho}\right)du
\qquad \,
\end{eqnarray}
with ${}_2F_1\left(\cdot,\cdot;\cdot;\cdot\right)$  denoting the
Gauss hypergeometric function [17, Eq. (9.100)]. Note that
(\ref{11}) and (\ref{12}) are valid only if $m_z$ is positive
integer \cite{12}. Using [18, Vol. 4, Eq. (3.35.7(4))] and the
Lauricella transformation to assure convergence [22, pp. 121],
(\ref{12}) is expressed in closed form as follows
\begin{eqnarray}\label{13}
W(k_1,\cdots,k_N) =
\Big(\prod_{j=1}^N\frac{\Gamma(m_z+k_j/2)}{\Gamma(m_z)}\Big)
\Big(\frac{1-\sqrt\rho}{1+(N-1)\sqrt\rho}\Big)^{m_z} \nonumber\\
\times \, F_A \left(m_z;\,m_z+\frac{k_1}{2},\cdots,m_z+\frac{k_N}{2};\,m_z,\cdots,m_z;\right. \qquad \quad \nonumber\\
\left. \frac{\sqrt\rho}{1 + (N-1)\sqrt\rho},\cdots
,\frac{\sqrt\rho}{1 + (N-1)\sqrt\rho}\right) \,, \qquad \,
\end{eqnarray}
where $F_A(\cdots)$   denotes the Lauricella hypergeometric
function of $N$ variables def\-ined by [17, Eq. (9.19)]. Note that
coeff\-icients $W(4)$, $W(2,2)$, $W(3,1)$   and $W(2,1,1)$ can be
expressed in terms of the more familiar hypergeometric functions
as per (B.1), (B.2), (B.4) and (B.6), respectively.

Introducing (\ref{8}) and (9) into (\ref{9}) and (11), one obtains the needed parameters
for the PDF approximation (\ref{7}) of $Z$ in closed form as
\begin{eqnarray}\label{14}
\Omega_R&=&\frac{E[Z^2]}{L} \,,\\
m_R&=&\frac{1+(L-1)\rho}{L}\frac{(E[Z^2])^2}{E[Z^4]-(E[Z^2])^2}
\,,
\end{eqnarray}
where $E[Z^2]$  and $E[Z^4]$  are respectively determined from (\ref{10}) and (\ref{11}).
Note that the fading parameter $m_R$  is typically calculated to a positive real number.

\subsection{Special Case: Sum of Identical Equally Correlated Nakagami RVs }

Let the equally correlated Nakagami RVs $Z_k$ ,$1\leq k\leq L$
have same second moments   $E[Z_k^2]=\Omega_Z$ (equipowered
branches), same fading parameter $m_Z$  (as positive integer) and
same correlation coeff\-icient $\rho$ between each pair of RVs. In
this case, (\ref{10}) and (\ref{11}) are simplif\-ied by using
(A.6) into
\begin{eqnarray}\label{15}
E[Z^2] = L\Omega_Z\left[1+\frac{(L-1)\Gamma^2(m_z+1/2)}{m_z \Gamma^2(m_z)}\right. \qquad \qquad \qquad \quad \nonumber\\
\times \, {}_2F_1\left(-\frac12,-\frac12;m_z;\rho\right)\bigg] \,, \qquad \quad \\
E[Z^4]=\left(\frac{\Omega_Z(1-\sqrt\rho)}{m_z}\right)^2
\, \Big[L \, W(4)+3L(L-1) \, W(2,2) \nonumber\\
+4 \, L(L-1) \, W(3,1) + 6L(L-1)(L-2) \, W(2,1,1) \nonumber\\
+ \, L(L-1)(L-2)(L-3) \, W(1,1,1,1)\Big] \,,  \qquad \qquad \,\,\,
\end{eqnarray}
where the necessary coeff\-icients $W(k_1,k_2,k_3,k_4)$  are again
calculated by (\ref{13}). The needed parameters for the PDF
approximation (\ref{7}) of $Z$ are obtained from (\ref{14}) and
(18).

\section{Application in the performance analysis of EGC receivers}
We consider a typical $L$-branch EGC diversity receiver exposed to
slow and f\-lat Nakagami fading. The envelopes of the useful
branch signals $Z_k$  are non-identical equally correlated
Nakagami random processes with PDFs given by (\ref{2}), whereas
their respective phases are i.i.d. uniform random processes. Each
branch is also corrupted by additive white Gaussian noise (AWGN)
with power spectral density $N_0/2$, which is added to the useful
branch signal. In the EGC receiver, the random phases of the
branch signals are compensated (co-phased), equally weighted and
then summed together to produce the decision variable.

The envelope of the composite useful signal, denoted by $Z$, is given by (\ref{1}),
whereas the composite noise power is given by $\sigma_{EGC}^2=LN_0/2$, resulting in the instantaneous output SNR given by
\begin{equation}\label{16}
\gamma_{EGC}=\frac{Z^2}{2\sigma_{EGC}^2}=\frac{1}{LN_0}\left(\sum_{k=1}^L
Z_k\right)^2=\left(\sum_{k=1}^L G_k\right)^2
\end{equation}
where RVs $G_k=Z_k/\sqrt{LN_0}$, $1\leq k\leq L$, form a set of
$L$ non-identical equally correlated Nakagami RVs with
$E[G_k^2]=\bar\gamma_k/L$, same fading parameters $m_z$ and same
correlation coeff\-icient $\rho$ among the diversity branches.
Note that $\bar\gamma_k=\Omega_k/N_0$  denotes the average SNR in
the $k$-th branch.

Using the results from Section II, it is now possible to
approximate PDF and MGF of (\ref{16}) by (\ref{6}) and (\ref{5}),
respectively, with $\Omega_R$  replaced by $\bar\gamma
=\Omega_R/(LN_0)$. These closed form approximations are then used
to determine the outage probability and the error probability of
$L$-branch EGC systems in correlated Nakagami fading with high
accuracy.
\subsection{Outage Probability}
The closed form approximation of the outage probability of the EGC
receiver (i.e. the CDF of $\gamma_{EGC}$) at threshold $t$ is
obtained by applying [18, Vol. 5, Eq. (2.1.3(1))] over (\ref{5})
as
\begin{eqnarray}\label{16a}
F_{\gamma_{EGC}}(t) \approx
\Big(\frac{m_Rt}{\bar{\gamma}(1+(L-1)\sqrt\rho)}\Big)^{m_R}
\Big(\frac{m_Rt}{\bar{\gamma}(1-\sqrt\rho)}\Big)^{m_R(L-1)}
\nonumber\\
\times \, \frac{1}{\Gamma(1+m_R L)} \,\, \Phi_2\Big(m_R,\,m_R(L-1);\,1+m_R L;  \qquad \qquad \nonumber\\
-\frac{m_R t}{\bar{\gamma}(1+(L-1)\sqrt\rho)}, \,-\frac{m_R
t}{\bar{\gamma}(1-\sqrt\rho)} \Big) \,, \qquad
\end{eqnarray}
where $\Phi_2(\cdot,\cdot;\cdot;\cdot,\cdot)$ denotes the
conf\-luent hypergeometric function of two variables def\-ined by
[17, Eq. (9.261(2))].

\subsection{Average Error Probability}
Comparing (\ref{1}) and (\ref{4}), it is obvious that the error
performance of an EGC system can be approximated by the
performance of an equivalent maximal ratio combining (MRC) system
for which many closed form solutions exist. For example, \cite{19}
derives the error probabilities of $L$-branch MRC with coherent
and non-coherent detection of binary signals in identical
correlated Nakagami fading channels. Thus, the average bit error
probabilities of the coherent BPSK system and non-coherent BFSK
are respectively expressed as [19, Eq. (32)] and [19, Eq. (26)]
\begin{eqnarray}\label{17}
\bar P_{C-BPSK}=\frac{1}{2}-\sqrt{\frac{1}{\pi}}\frac{\Gamma(m_R
L+1/2)}{\Gamma(m_R L)}
\left[\frac{\bar\gamma(1-\sqrt\rho)}{m_R+\bar\gamma(1-\sqrt\rho)}\right]^{\frac{1}{2}}\nonumber\\
\times\left[\frac{m_R}{m_R+\bar\gamma(1-\sqrt\rho)}\right]^{m_R L}
\left[\frac{1-\sqrt\rho}{1+(L-1)\sqrt\rho}\right]^{m_R} \qquad \qquad \nonumber\\
\times \, F_2\left(m_R L+\frac{1}{2};1,m_R;\frac{3}{2},m_R L
;\frac{\bar\gamma(\sqrt\rho-1)}{m_R+\bar\gamma(1-\sqrt\rho)}\right., \qquad \nonumber\\
\frac{m_R L \sqrt\rho}
{(m_R+\bar\gamma(1-\sqrt\rho))(1+(L-1)\sqrt\rho} \bigg) \, ,
\qquad
\end{eqnarray}
and
\begin{equation}\label{18}
\bar P_{NC-BFSK}=\frac{1}{2} \, M_{R^2}(s)\bigg|_{s=-\frac{1}{2}}
\,. \qquad \qquad \qquad \qquad \quad
\end{equation}
where $F_2(\cdot;\cdot,\cdot;\cdot,\cdot;\cdot,\cdot)$  is the
Appell hypergeometric function (as the special case of Lauricella
$F_A$ function of two variables) def\-ined by [17, Eq.
(9.180(2))].

\section{ Illustrative Examples and Discussion }
In this Section, the proposed approximation for the sum of arbitrary number of non-identical equally correlated
Nakagami channels is validated by Monte-Carlo simulations. The simulation of correlated Nakagami random signals
is realized by using the method proposed in [21, Section VII].

In order to model the non-identical branch signals (i.e., unequal
average branch powers and unequal average branch SNRs), we
introduce exponentially decaying prof\-ile, modeled as
\begin{equation}\label{19}
\Omega_k=\Omega_1\exp(-\delta(k-1)),\qquad 1\leq k\leq L,
\end{equation}
where $\Omega_1$  is the average power of branch 1 ($k = 1$) and $\delta$  is the decaying exponent.
Note that $\delta=0$  denotes the case of identical branch signals (i.e., equal branch powers and equal average branch SNRs).

Fig. 1 illustrate the high accuracy of the proposed PDF
approximation of RV (\ref{1}) for a large variety of fading
scenarios.

Figs. 2 and 3 illustrate the high accuracy of the equivalent BPSK
MRC error probability (\ref{17}) for evaluation of the
approximated BPSK EGC error probability.

\begin{figure}
\centering
\includegraphics[width=3.5in]{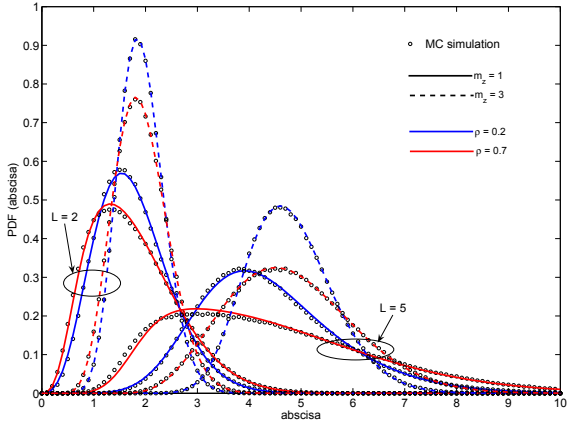}
\vspace{-5mm} \caption {PDF approximation of a sum of $L$
identical equally correlated Nakagami RVs for $\Omega_1 = 1$ and
$\delta = 0$} \label{fig_1} \vspace{-3mm}
\end{figure}


\useRomanappendicesfalse
\appendices
\section{}
\renewcommand{\theequation}{\thesection.\arabic{equation}}
\setcounter{equation}{0} In order to determine $E[Z^2]$ and
$E[Z^4]$, we apply the multinomial theorem [20, Eq. (24.1.2)]. The
second moment $E[Z^2]$ can be extracted straightforwardly. The
fourth moment $E[Z^4]$, after using [4, Eq. (43)] and performing
some mathematical manipulations, can be transformed to
\begin{eqnarray*}\label{23a}
E[Z^4]&=&\left(\frac{1-\sqrt\rho}{m_z}\right)^2\bigg[W(4)
\sum_{k=1}^L\Omega_k^2+6 W(2,2)\nonumber\\
\end{eqnarray*}
\begin{eqnarray*}\label{23a}
&\times&\sum_{i=1}^L\sum_{j=i+1}^L\Omega_i\Omega_j +4 W(3,1)
\sum_{i=1}^L\sum_{j=i+1}^L\sqrt{\Omega_i^3\Omega_j}\nonumber\\
&+&4 W(1,3)
\sum_{i=1}^L\sum_{j=i+1}^L\sqrt{\Omega_i\Omega_j^3}\nonumber\\
&+&12 W(2,1,1)
\sum_{m=1}^L\sum_{i=m+1}^L\sum_{j=i+1}^L\sqrt{\Omega_m^2\Omega_i\Omega_j}\nonumber\\
&+&12 W(1,2,1)\sum_{m=1}^L\sum_{i=m+1}^L\sum_{j=i+1}^L\sqrt{\Omega_m\Omega_i^2\Omega_j}\nonumber\\
&+&12 W(1,1,2)\sum_{m=1}^L\sum_{i=m+1}^L\sum_{j=i+1}^L
\sqrt{\Omega_m\Omega_i\Omega_j^2}\nonumber\\
\end{eqnarray*}
\vspace{-0.7cm}
\begin{equation}
+24W(1,1,1,1)\sum_{m=1}^L\sum_{n=m+1}^L\sum_{i=n+1}^L\sum_{j=i+1}^L
\sqrt{\Omega_n\Omega_m\Omega_i\Omega_j}\bigg]
\end{equation}
It is obvious from (\ref{12}) that $W(3,1)=W(1,3)$ and $W(2,1,1) =
W(1,2,1) = W(1,1,2)$, which directly yields to the result given by
(\ref{11}).

\section{}
\setcounter{equation}{0} Using identities given by [20, Eqs.
(13.6.9) and (22.3.9)], ${}_1F_1(0,m_z,-au)=1$,
${}_1F_1(-1,m_z,-au)=1+au/m_z$ and
${}_1F_1(-2,m_z,-au)=1+2au/m_z+(au)^2/(m_z(1+m_z))$, one directly
obtains
\begin{eqnarray}\label{24}
W(4)=\left(\frac{\Gamma(m_z+2)}{\Gamma(m_z)}\right)
\frac{1}{\Gamma(m_z)} \qquad \qquad \qquad \qquad \qquad \nonumber\\
\times\int_0^\infty u^{m_z-1}e^{-u}
{}_1F_1\left(-2,m_z;-au\right)du \qquad \nonumber\\
=m_z(1+m_z)(1+a)^2 \,, \qquad \qquad \qquad \qquad \qquad
\end{eqnarray}
\begin{eqnarray}\label{25}
W(2,2)=\left(\frac{\Gamma(m_z+1)}{\Gamma(m_z)}\right)^2
\frac{1}{\Gamma(m_z)} \qquad \qquad \qquad \qquad \quad \nonumber\\
\times \int_0^\infty u^{m_z-1}e^{-u}
\big[{}_1F_1\left(-1,m_z;-au\right)\big]^2 du \qquad \quad \nonumber\\
 = a^2m_z+m_z^2(1+a)^2 \,, \qquad \qquad \qquad \qquad \qquad
\end{eqnarray}
where $a=\sqrt\rho/(1-\sqrt\rho)>0$. Using [17, Eqs. (9.212 (1))
and (7.622 (1))], it is possible to obtain the following identity
\begin{eqnarray}\label{26}
J(m,a,p,q)= \frac{1}{\Gamma(m)} \int_0^\infty u^{m-1}e^{-u}
{}_1F_1\left(-\frac{p}{2},m;-au\right)\nonumber\\
\times {}_1F_1\left(-\frac{q}{2},m;-au\right)du =
(1+a)^\frac{p}{2} \left(\frac{1+2a}{1+a}\right)^\frac{q}{2}
\nonumber\\
\times \,
{}_2F_1\left(m+\frac{p}{2};-\frac{q}{2};m,-\frac{a^2}{1+2a}\right)
\,, \qquad
\end{eqnarray}
resulting into
\begin{eqnarray}\label{27}
W(3,1)=\left(\frac{\Gamma(m_z+3/2)}{\Gamma(m_z)}\frac{\Gamma(m_z+1/2)}{\Gamma(m_z)}\right)
\frac{1}{\Gamma(m_z)} \qquad \qquad \quad \nonumber\\
\int_0^\infty u^{m_z-1}e^{-u}
{}_1F_1\left(-\frac{3}{2},m_z;-au\right)
{}_1F_1\left(-\frac{1}{2},m_z;-au\right)du \nonumber\\
=\frac{\Gamma(m_z+3/2)}{\Gamma(m_z)}\frac{\Gamma(m_z+1/2)}{\Gamma(m_z)}
J(m_z,a,3,1) \,. \qquad \qquad
\end{eqnarray}
Using [17, Eq. (9.212 (3)), pp. 1023] and after some simple
algebra, one obtains the following
\begin{eqnarray*}\label{29}
\left[{}_1F_1\left(-\frac{1}{2},m;-au\right)\right]^2 \qquad \qquad \qquad \qquad \qquad \qquad \qquad \qquad \quad \nonumber\\
=\left(\frac{m+1/2}{m}\right)^2\left[{}_1F_1\left(-\frac{1}{2},m+1;-au\right)\right]^2 \qquad \qquad \qquad \qquad  \nonumber\\
+\frac{1}{4m^2}\left[{}_1F_1\left(\frac{1}{2},m+1;-au\right)\right]^2 \qquad \qquad \qquad \qquad \qquad \qquad \nonumber\\
-\frac{m+1/2}{m^2}{}_1F_1\left(\frac{1}{2},m+1;-au\right){}_1F_1\left(-\frac{1}{2},m+1;-au\right)
\qquad
\end{eqnarray*}
\vspace{-0.5cm}
\begin{equation}
\end{equation}
Thus,
\begin{eqnarray*}\label{30}
W(2,1,1)=\left(\frac{\Gamma(m_z+1)}{\Gamma(m_z)}\left(\frac{\Gamma(m_z+1/2)}{\Gamma(m_z)}\right)^2\right)
\frac{1}{\Gamma(m_z)} \qquad \qquad \nonumber\\
\int_0^\infty u^{m_z-1}e^{-u} {}_1F_1\left(-1,m_z;-au\right)
\left[{}_1F_1\left(-\frac{1}{2},m_z;-au\right)\right]^2du \quad \nonumber\\
=m_z\left(\frac{\Gamma(m_z+1/2)}{\Gamma(m_z)}\right)^2\bigg[J(m_z,a,1,1)
+\frac{a(m_z+1/2)^2}{m_z^2} \qquad  \nonumber\\
\times \, J(m_z+1,a,1,1) +\frac{a}{4 m_z^2} J(m_z+1,a,-1,-1) \qquad \,\,  \nonumber\\
-\frac{a(m_z+1/2)}{m_z^2}J(m_z+1,a,-1,1)\bigg] \qquad
\end{eqnarray*}
\begin{equation}
\end{equation}


\begin{figure}
\centering
\includegraphics[width=3.5in]{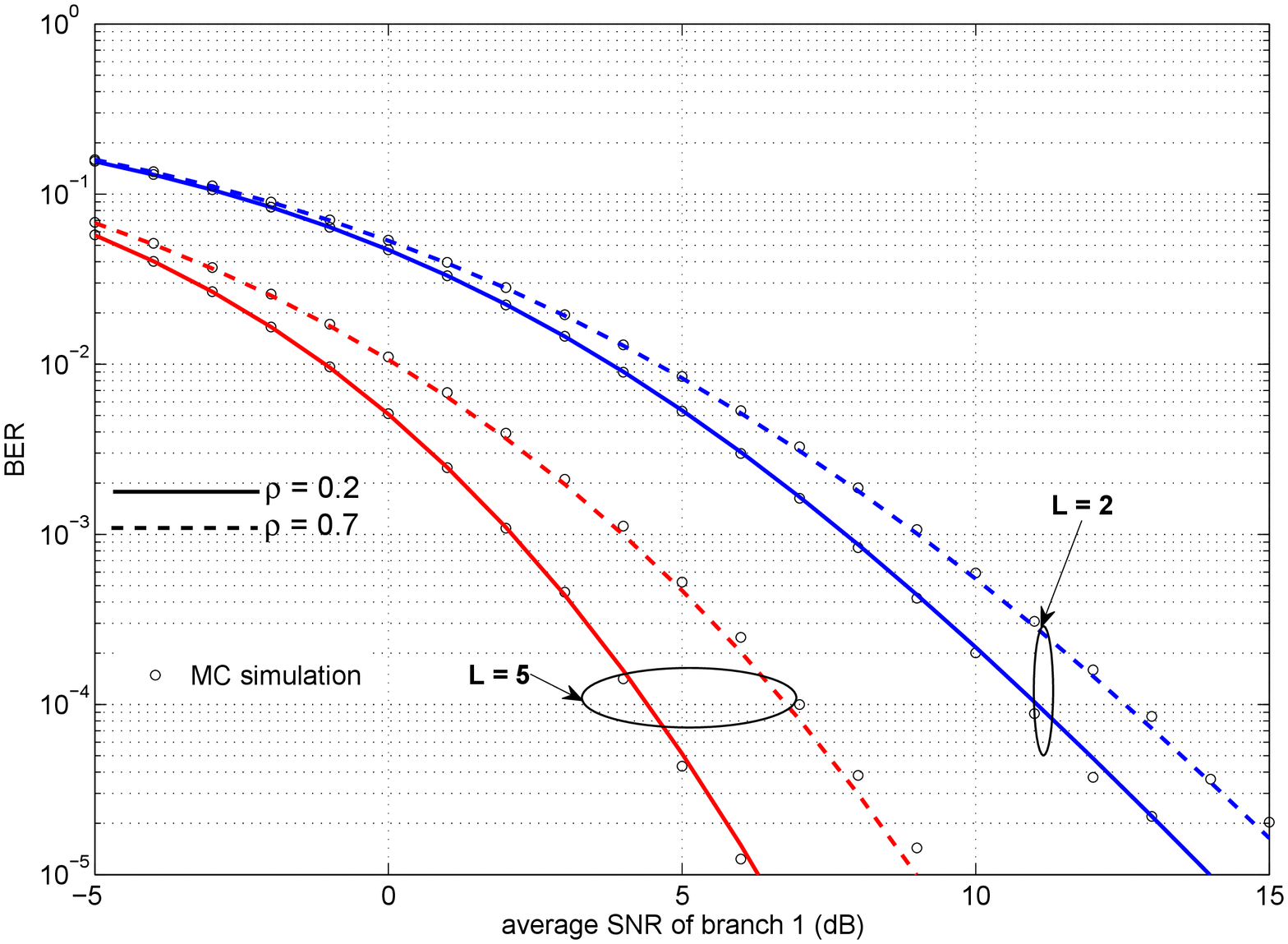}
\vspace{-5mm} \caption {BER of equivalent MRC \textbf{BPSK} system
for $m_z = 2$ and $\delta = 0$} \label{fig_2} \vspace{-3mm}
\end{figure}


\begin{figure}
\centering
\includegraphics[width=3.5in]{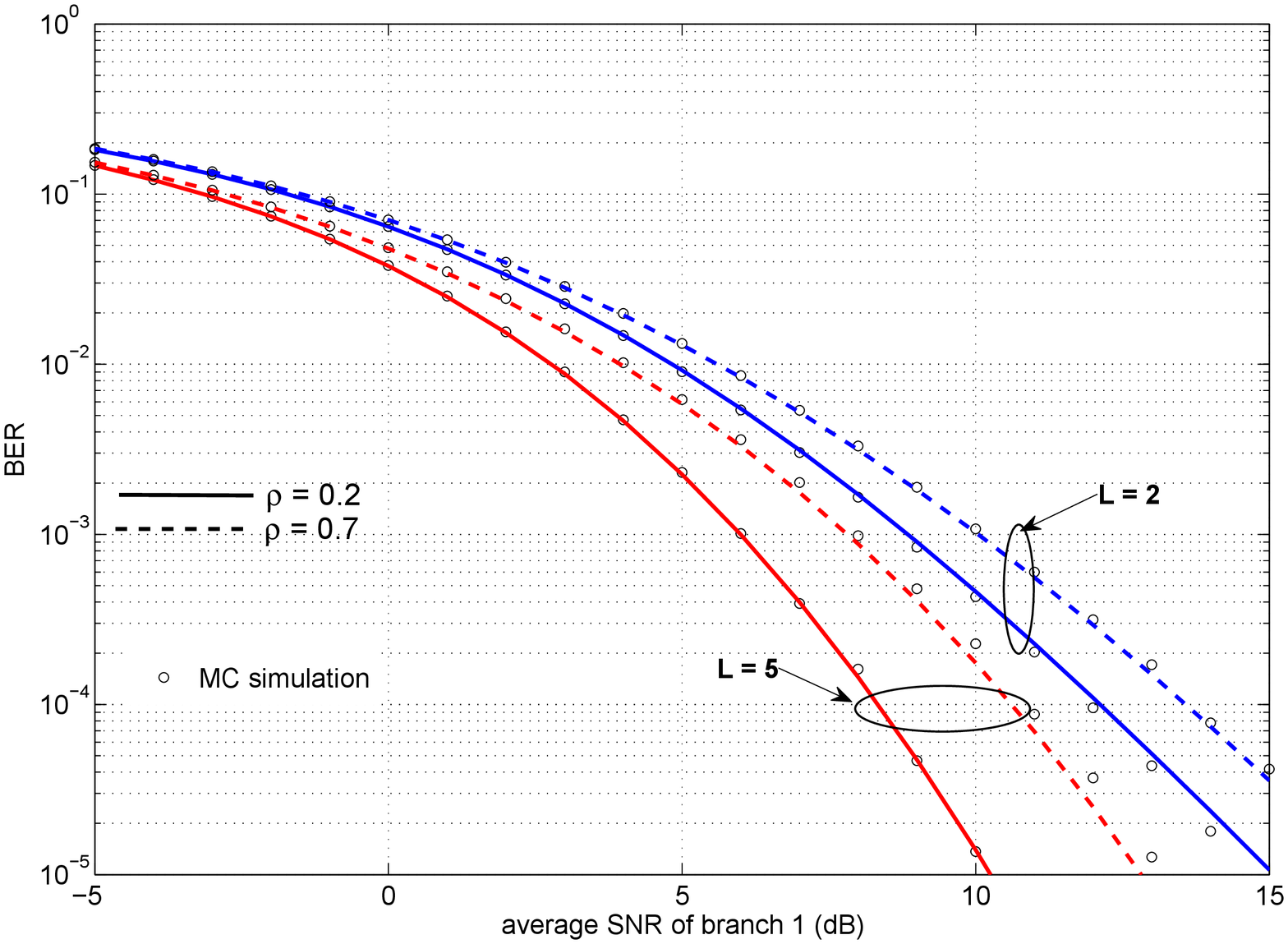}
\vspace{-5mm} \caption {BER of equivalent MRC \textbf{BPSK} system
for $m_z = 2$ and $\delta = 0.5$} \label{fig_2} \vspace{-3mm}
\end{figure}







\end{document}